# A Preliminary Analysis on the Effects of Propensity to Trust in Distributed Software Development

Fabio Calefato, Filippo Lanubile, Nicole Novielli
University of Bari
Bari, Italy
{fabio.calefato | filippo.lanubile | nicole.novielli}@uniba.it

*Abstract*—**Establishing trust between developers working at distant sites facilitates team collaboration in distributed software development. While previous research has focused on how to build and spread trust in absence of direct, face-to-face communication, it has overlooked the effects of the propensity to trust, i.e., the trait of personality representing the individual disposition to perceive the others as trustworthy. In this study, we present a preliminary, quantitative analysis on how the propensity to trust affects the success of collaborations in a distributed project, where the success is represented by pull requests whose code changes and contributions are successfully merged into the project's repository.**

## I. INTRODUCTION

Trust is critical to the success of global software engineering projects. Reduced trust has been reported to (a) aggravate the feeling of being separate teams with conflicting goals, (b) decrease the willingness to share information and cooperate to solve problems, and (c) affect goodwill toward others in case of objections and disagreements [1]. Trust among project members typically grows through close face-to-face (F2F) interaction, as it represents the most effective way to establish connections with others and gain awareness of both technical and personal aspects [2],[3],[4]. Unfortunately, F2F interaction is also the very activity that is largely reduced in distributed software projects, or even completely unavailable. Prior empirical research has also found that engaging in social interactions over email [5] or chat [6],[7] has a trust-building effect among members of open-source software (OSS) projects who typically have no chances to meet.

An essential aspect for understanding the development of trust and cooperation in a team is the *propensity to trust*, that is, the varying, personal disposition of the trustor to 'take the risk' intrinsically associated with believing that the trustee(s) will behave as expected [8]. In other words, *propensity to trust* refers to an individual's general tendency to perceive the other individuals as trustworthy [9]. We formulate our research question as:

*RQ: How does individual propensity to trust facilitate successful collaborations in globally-distributed software projects?*

One common limitation identified in prior empirical-research findings on trust is that there is no explicit measure of the extent to which developers' trust contributes to project performance. In this study, we intend to overcome this limitation by approximating the overall performance of a project (e.g., requirements completion, productivity, duration) with the history of successful collaborations occurring between project developers. By successful collaborations, we indicate situations where (at least) two developers work together and their cooperation is successful because it yields a project advancement (e.g., by fixing bugs or adding new features). Through such a fine-grain unit of analysis, we aim to measure 'more directly' how *trust* facilitates cooperation.

Modern distributed software projects support their workflow and coordinate remote work through version control systems. A pull request is a popular way of submitting contributions to a project using a distributed version control system such as Git. According to the pull-based development model [10], the project's main repository is not shared among developers. Instead, developers contribute by forking (i.e., cloning) the repository and making their changes independently from each other. When a set of changes is ready to be submitted to the main repository, a potential *contributor* creates a pull request. Then, an *integration manager,* one of the core developers, is assigned the responsibility to inspect the changes and integrate them into the project's main development line. The role of the integration manager is crucial to ensure project quality. After a contribution has been received, the integrators must close a pull request deciding whether it is suitable for the project – i.e., the pull request is either *accepted* and changes are *integrated* into the project's main repository – or considered incorrect – i.e., the pull request is *declined* and changes are *rejected*. Closed pull requests, whether accepted or declined, require that a consensus is reached through discussion. Collaborative development platforms such as GitHub and Bitbucket [11], make it easier for developers to collaborate through pull requests as they provide a user-friendly web interface for discussing proposed changes before integrating them into the project source code.

Accordingly, we represent successful collaborations between developers in terms of accepted pull requests and refine our research question as follows:

*RQ': How does individual propensity to trust facilitate the acceptance of pull-requests in globally-distributed software projects?*

We investigate the refined research question by analyzing the history of contributions (pull requests) from the developers of Apache Groovy. Since Groovy provides the archived history of email-based communications, we analyze the interaction traces over such channel to assess the developers' *propensity to trust*.

The remainder of the paper is organized as follows. In Section 2, we discuss the challenges of and solutions to quantifying the propensity to trust. In Section 3 and 4, respectively, we describe the empirical study and its results. In Section 5 we discuss findings and limitations. Finally, we draw conclusions and describe future work in Section 6.

## II. BACKGROUND

### A. Measuring Propensity to Trust

The *Big-Five personality model* (or *Five-Factor model*) [12], is a general taxonomy of personality traits that includes, at the higher level: *openness*, *conscientiousness*, *extraversion*, *agreeableness*, and *neuroticism* (see Figure 1). Each top-level dimension has six sub-dimensions, or facets, that further characterize an individual according to the dimension. Previous research has confirmed that the personality traits can be successfully derived from the analysis of written text [13] such as emails [14]. In fact, Tausczik & Pennebaker [15] found that every trait in the Big-Five model is strongly and significantly associated with theoretically-appropriate patterns of word use, indicating strong connections between language use and personality.

Previous research on trust has relied on self-reported data, typically survey questionnaires, to measure individual's trust on a given scale [16],[17],[18],[19]. One notable exception is represented by the work of Wang & Redmiles [7], who studied how trust spreads in OSS projects. They used the Linguistic Inquiry and Word Count (*LIWC*) psycholinguistics dictionary to analyze word usage in writing [15],[20].

To obtain a quantitative measure of trust, we relied on *Tone Analyzer*,[1] an IBM Watson service leveraging LIWC, which uses linguistic analysis to detect three types of tones from written text: social, emotional, and writing style. Specifically, the *social tone* measures the social tendencies (i.e., the Big-Five personality traits) in people's writing. In particular, we focused on the recognition of *agreeableness*, the personality trait indicating a person's tendency to be compassionate and cooperative toward others. One facet of agreeableness is the tendency to trust the others rather than being suspicious [21]. Accordingly, in the following, we use **agreeableness as a proxy measure of the individual's propensity to trust**.

### B. Factors influencing pull-requests acceptance

The factors that influence the acceptance of contributions are both social and technical in nature [22],[23].

Regarding the technical aspects, previous research on patch acceptance [24], code reviewing [25], and bug triaging [26] has found that the decision to merge contributions is affected by size of both the project (i.e., KLOC and team size) and the patch itself (i.e., the number of files changed), the presence of test cases, and the extent to which changes are discussed (i.e., number of comments and participants in the review). However, Gousios et al. [10] found that only ~13% of the reviewed pull requests were closed without merging for purely technical reasons. In particular, they found the decision to merge to be mostly affected by whether the changes involved an area of code actively under development and the coverage of the attached test cases.

With the raise of 'transparent' social-coding platform such as Bitbucket and GitHub, integrators make inferences about the quality of contributions, not only by looking at their technical quality but also using developer's track record (e.g., previous contributions accepted) and reputation (e.g., number of stars and followers in GitHub) as auxiliary indicators [27],[28]. An interesting finding is that pull requests are 'treated equally' regardless of the submitters' 'social status,' i.e., whether they are external contributors rather than members of the core development team [10],[29]. Ducheneaut [30] found that contributions coming from submitters who are known to the core development team have higher chances of being accepted, as core developers also use the record of interactions as signals for judging the quality of proposed changes.

Overall, these findings provide further motivation to our work for **looking at other, non-technical factors that may influence the decision to merge pull requests**.

## III. EMPIRICAL STUDY

We designed a study to quantitatively assess the impact of the propensity to trust to pull requests (PRs) acceptance. We used a simple logistic regression to build a model for estimating the probability of 'success' of a pull request (i.e., *merged*) given the integrator's *propensity to trust* (i.e., its *agreeableness* as measured by the IBM Watson Tone Analyzer). Therefore, in our framework, we treat the acceptance of a pull request as the dependent variable while the measured agreeableness of the integrator is the independent variable (i.e., the predictor).

In this study, the two sources of information used to collect data are the pull requests in GitHub and the emails retrieved from the Apache Groovy project (see Table 1), an object-oriented programming and scripting language for the Java platform. Among the many projects that are supported by the Apache Software Foundation, we opportunistically selected Groovy because a) it makes its mailing list archives freely accessible and b) it follows a pull request-based development model.

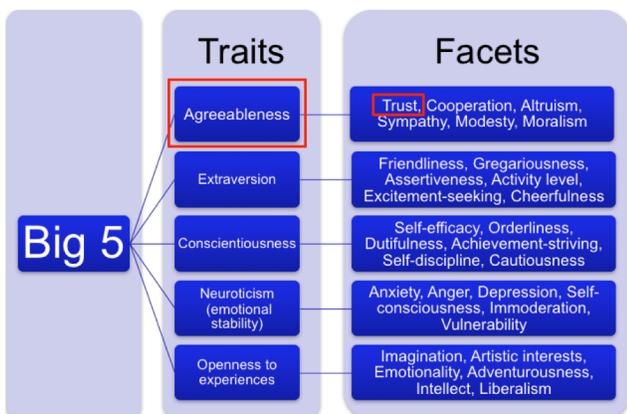

Fig 1. Big-five personality traits model.

---
[1] www.ibm.com/watson/developercloud/tone-analyzer.html

*A. Dataset*

We used the *GHTorrent* [31] database to retrieve the chronologically-ordered list (i.e., history) of pull requests opened on GitHub between March 2015 and December 2016. For each pull request, in particular, we stored the following information (i) the *contributor*, (ii) the *date when it was opened*, (iii) the *status* (i.e., *merged* or *not*), (iv) the *integrator* who merged it, if any, and (v) the *date when it was it closed or merged*.

Not all the pull requests are merged through GitHub, though. To identify those that were closed and merged outside of GitHub, we looked at the pull-request comments (see Figure 2). In particular, we searched for the presence of a) commits to the master branch that closed the pull request and b) comments from the integration manager who acknowledge the successful merge. All the project's pull requests were reviewed by one of the researchers and their status manually annotated. Albeit entirely automated, a similar procedure is described by Gousios et al. [10].

Regarding the project's emails, we retrieved almost 5,000 messages. We used the *mlstats*[2] tool to mine the user and dev mailing-list archives available on the Groovy project website.[3] To do so, we first retrieved the identities of the committers (i.e., core-team members with write access to the repository) from the web page of the Groovy project hosted at both GitHub[4] and Apache.org[5]). Then, we compare ids and names of the pull-request integrators against the names and email addresses of the senders of messages shared on the project mailing lists. We were able to identify the messages from 10 integrators. Finally, we filtered out those developers who had exchanged less than 20 email messages between Mar. 2015 and Dec. 2016 (i.e., less than one email per month). This preprocessing step was necessary to ensure that a significant amount of non-technical content from each developer was available for the analysis and to measure the approximate levels of *propensity to trust*. Eventually, we ended up selecting 6 integrators who had reviewed 218 pull requests.

*B. Integrators' Propensity to Trust*

Once we obtained a mapping of the core-team members and their communication records, we computed the *propensity to trust* scores from the content of the entire corpus of their emails.

In particular, we processed the content of the emails using Tone Analyzer and obtained an agreeableness score, which is defined within the interval [0, 1]. Values smaller than 0.5 are associated with *low agreeableness* and therefore, to the tendency to be less compassionate and cooperative towards others. Instead, values equal to or greater than 0.5 are in general associated with *high agreeableness*. Accordingly, albeit computed as numeric, we transformed the score into a nominal variable with two levels, namely {*Low*, *High*}, Thus, at the end of the process, for each core-team member we obtained a *Low/High* agreeableness score, which we consider the approximate level of an integrator's *propensity to trust* (see Table 2).

## IV. RESULTS

In this section, we present the results of the regression model built to understand whether *propensity to trust* is a predictor of pull-request acceptance. In particular, we performed a simple logistic regression using the R statistical package.

The results of the analysis are reported in Table 3, where we omit to report the positive and significant effect of control variables `#PRs reviewed` and `#emails sent` due to space constraints. Regarding the predictor `Propensity to trust`, we note that the coefficient estimate (+1.49), the odds ratio (4.46), and the statistical significance (p-value = 0.009). The sign of the coefficient estimate indicates the positive/negative association of the predictor with the success of a pull request. The odds ratio (OR) weighs the effect size of this impact: the closer its value to 1, the smaller the impact of the parameter on the chance of success. A value lower than 1 corresponds to a negative association of the predictor (negative sign of the coefficient estimate) with success, and the opposite

TABLE I. DESCRIPTION OF APACHE GROOVY PROJECT.

| Description | Object-oriented programming language for the Java platform |
|---|---|
| Language | Java |
| # project committers* | 12 |
| # PRs on GitHub* | 476 |
| # emails archived in mailing-list* | 4948 |
| # unique email senders† | 367 |

\* As of Jan. 2017
† Multiple email addresses from the same developer counted as 1

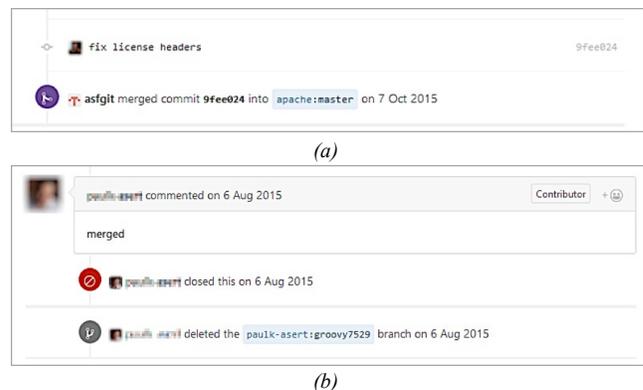

*(a)*

*(b)*

Fig 2. Two examples of pull requests: one merged in GitHub (a), the other one merged outside GitHub and closed (b).

---

[2] https://github.com/MetricsGrimoire/MailingListStats
[3] http://groovy-lang.org/mailing-lists.html
[4] https://github.com/apache/groovy
[5] http://incubator.apache.org/projects/groovy.html

for a value higher than 1. An OR=*x* technically implies that the odds of the positive outcome are *x* times greater than the odds of the negative outcome [32]. Accordingly, these results indicate that the *propensity to trust* is positively and significantly associated with the probability of a pull request to be successfully merged (p<0.001).

Furthermore, we can use the coefficients estimates from our model (see Table 2) in the equation (1) in R[6] to make sense of the extent to which the propensity to trust of different integrators can affect the probability of merging a given pull request from the Groovy project.

$$Estimated\ Prob. of\ PR\ acceptance = 1/\left(1 + exp\left(-(+1.49 + 4.46 * Propensity)\right)\right) \quad (1)$$

For example, given the pull request $k$, the probability of being accepted by an integrator $i$ with a *Low propensity to trust* is $p_i^k = 0.68$. If the same pull request $k$ were to be reviewed by another developer $j$ with a *High propensity to trust*, the probability of being merged would be $p_j^k = 0.91$, which corresponds to an increase of ~34%.

## V. DISCUSSION

To the best of our knowledge, this is the first attempt at quantifying the effects that trust or other developers' personal traits have on software projects following a pull request-based development model. As such, the main result of this study is the initial evidence that the chances of merging code contributions are correlated with the personality traits of the integrators who perform the code review. On one hand, this is a novel finding that underlines the role played by developers' propensity to trust, and more generally by personality, in the execution of complex tasks such as code reviewing. On the other hand, this result is in line with the finding of Tsay et al. [22] and Ducheneaut [30] who observed that the social distance between the contributor and the integrator also influence the chances of accepting a pull request.

TABLE II. PROPENSITY TO TRUST SCORES AND PULL REQUESTS MERGED FOR THE GROOVY PROJECT INTEGRATORS.

| Integrator | PR reviewed | | Propensity to Trust score |
|---|---|---|---|
| | merged | closed | |
| Dev 1 | 57 | 7 | High |
| Dev 2 | 10 | 0 | High |
| Dev 3 | 99 | 6 | High |
| Dev 4 | 12 | 1 | Low |
| Dev 5 | 8 | 4 | Low |
| Dev 6 | 14 | 0 | Low |
| Tot. | 200 | 18 | -- |

TABLE III. RESULTS OF THE SIMPLE LOGISTIC REGRESSION.

| Predictor | Coefficient estimate | Odds ratio | p-value |
|---|---|---|---|
| (Intercept) | +0.77 | -- | 0.117 |
| Propensity to trust | +1.49 | 4.46 | 0.009* |

\* significant at 0.01 level

---

[6] https://www.r-project.org/

Accordingly, they recommended developers to make sure to be known in the community before sending contributions. Given our finding, a follow-up to this recommendation would be to also use `@user` mentions in the pull request comments to explicitly request the review from integrators who have shown more willingness to cooperate and help others before (i.e., higher *propensity to trust*).

Furthermore, previous research has found that, unlike patches, most pull requests are successfully merged. In fact, while Baysal et al. [33] found that only about a half of the submitted patches to the Chrome and Firefox projects made it into the repository, Gousios et al. [10] found instead that about 80% of the ~170k pull requests analyzed in their study were merged. Our result finding is in line with this finding since we observed that the probability of getting a PR accepted is above chance (p=0.68), even in the case of integrators with a low *propensity to trust*. Nonetheless, it is worth noting that the chances of success increase by more than 30% (p=0.91) if the integrator in charge of reviewing the contribution exhibits instead a high *propensity to trust*.

Overall, in the broader framework of sociotechnical congruence research, our finding calls for further studies to investigate the so-far-neglected effects of trust and other personality traits when matching the coordination needs established by the technical domain (e.g., the area of the source code interested by the proposed changes) and the actual coordination activities carried out by the development team (i.e., assigning the code review tasks to one of the core developers).

Finally, mainly due to its preliminary nature, this study suffers from some limitations. Regarding the generalizability of results, we acknowledge that the analysis is based on a limited number of pull requests involving the developers from a single project. Only through replications with different settings and larger datasets we will be able to develop a more solid empirical evidence.

The other limitations revolve around the validity of the *propensity to trust* construct. Because of their lack of practicality, we decided not to rely on traditional, self-reported psychometric approaches for measuring trust (e.g., surveys). Instead, we relied on the IBM Watson Tone Analyzer to approximate the level the of *propensity to trust* in term of *agreeableness*, a personality trait is associated with a tendency to trust and cooperate with others [12]. In future replications, we will investigate the reliability of the Tone Analyzer service also in a technical domain such as software engineering, since linguistic resources are usually trained on non-technical content.

## VI. CONCLUSIONS & FUTURE WORK

This work represents the first step in a broader research effort to collect quantitative evidence that establishing trust among developers contribute to project performance. According to the Big-Five personality model, the propensity to perceive others as trustworthy is a stable personality trait that varies with individuals. Thus, leveraging prior evidence that personality emerges unconsciously from the personal lexicon used in written communication, we used the IBM Watson Tone

Analyzer service to measure the *propensity to trust* through the analysis of the emails archived by Apache Groovy project. We found initial evidence that the developers with higher propensity to trust are more likely to accept external contributions in form of pull requests.

As future work, we intend to replicate the experiment to obtain more solid evidence. Specifically, we intend to compare the Tone Analyzer with other tools to better assess their reliability in extracting personality from text containing technical content. Furthermore, we intend to enlarge the dataset in terms of both projects and pull requests in order to understand whether and how: a) developers' personality changes depending on the project they participate in; b) mutual trust evolves over time between pairs of developers who interact in a dyadic cooperation.

ACKNOWLEDGMENT

We are grateful to IBM for providing free access to the Bluemix platform. This work is partially funded by the project 'Investigating the Role of Emotions in Online Question & Answer Sites' under the program "Scientific Independence of young Researchers" (SIR), funded by the Italian Ministry of University and Research (MIUR).